\documentclass [12pt]{article}
\usepackage{graphicx,amssymb,amsfonts,latexsym,amsmath,amsthm,times}
\usepackage{epsfig}
\usepackage{fancyhdr}
\usepackage{color}
\usepackage[english]{babel}
\numberwithin{equation}{section}

\newtheorem{theorem}{Theorem}

\newtheorem{remark}[theorem]{Remark}

\renewcommand\thesection{\arabic{section}.}
    \renewcommand\thesubsection{\thesection\arabic{subsection}.}

\begin{document}

\footnotesize {\flushleft \mbox{\bf \textit{Math. Model. Nat.
Phenom.}}}
 \\
\mbox{\textit{{\bf Vol. 8, No. 6, 2010, pp. 1-21}}}

\thispagestyle{plain}

\vspace*{2cm} \normalsize \centerline{\Large \bf Spatiotemporal dynamics in a spatial plankton system}

\vspace*{1cm}

\centerline{\bf Ranjit Kumar Upadhyay\footnote{Corresponding
author. E-mail: ranjit\_ism@yahoo.com. Present Address is:
Institute of Biology, Department of Plant Taxonomy and Ecology, Research group of Theoretical Biology and Ecology, E$\ddot{o}$tv$\ddot{o}$s Lorand  University, H-1117, Pazmany P.S. 1/A, Budapest, Hungary.
}$^a$,  Weiming Wang\footnote{E-mail: weimingwang2003@163.com}$^{b,c}$
 and N. K. Thakur$^a$ }

\vspace*{0.5cm}

\centerline{$^a$ Department of Applied Mathematics,
Indian School of Mines, Dhanbad 826004, India}

\centerline{$^b$ School of Mathematical Sciences, Fudan University, Shanghai, 200433 P.R. China}

\centerline{$^c$ School of Mathematics and Information Science,
Wenzhou University,} \centerline{Wenzhou, Zhejiang, 325035 P.R.China}


\vspace*{1cm}

\noindent {\bf Abstract.}
In this paper, we investigate the complex dynamics of a spatial plankton-fish system with Holling type III functional responses. We have carried out the analytical study for both one and two dimensional system in details and found out a condition for diffusive instability of a locally stable equilibrium. Furthermore, we present a theoretical analysis of processes of pattern formation that involves organism distribution and their interaction of spatially distributed population with local diffusion. The results of numerical simulations reveal that, on increasing the value of the fish predation rates, the sequences spots $\rightarrow$ spot-stripe mixtures$\rightarrow$ stripes$\rightarrow$ hole-stripe mixtures holes$\rightarrow$ wave pattern is observed. Our study shows that the spatially extended model system has not only more complex dynamic patterns in the space, but also has spiral waves.

\vspace*{0.5cm}




\vspace*{1cm}

\setcounter{equation}{0}
\section{Introduction}

Complexity is a tremendous challenge in the field of marine ecology. It impacts the development of theory, the conduct of field studies, and the practical application of ecological knowledge. It is encountered at all scales (Loehle, 2004). There is growing concern over the excessive and unsustainable exploitation of marine resources due to overfishing, climate change, petroleum development, long-range pollution, radioactive contamination, aquaculture etc. Ecologists, marine biologists and now economists are searching for possible solutions to the problem (Kar and Matsuda, 2007). A qualitative and quantitative study of the interaction of different species is important for the management of fisheries. From the study of several fisheries models, it was observed that the type of functional response greatly affect the model predictions (Steele and Henderson, 1992; Gao et al., 2000; Fu et al., 2001; Upadhyay et al., 2009). An inappropriate selection of functional response alters quantitative predications of the model results in wrong conclusions. Mathematical models of plankton dynamics are sensitive to the choice of the type of predator's functional response, i.e., how the rate of intake of food varies with the food density. Based on real data obtained from expeditions in the Barents Sea in 2003-2005, show that the rate of average intake of algae by Calanus glacialis exhibits a Holling type III functional response, instead of responses of Holling types I and II found previously in the laboratory experiments. Thus the type of feeding response for a zooplankton obtained from laboratory analysis can be too simplistic and misleading (Morozov et al., 2008). Holling type III functional response is used when one wishes to stabilize the system at low algal density (Truscott and Brindley, 1994; Scheffer and De Boer, 1996; Bazykin, 1998; Hammer and Pitchford, 2005).

Modeling of phytoplankton-zooplankton interaction takes into account zooplankton grazing with saturating functional response to phytoplankton abundance called Michaelis-Menten models of enzyme kinetics (Michaelis and Menten, 1913). These models can explain the phytoplankton and zooplankton oscillations and monotonous relaxation to one of the possible multiple equilibria (Steele and Henderson, 1981, 1992; Scheffer, 1998; Malchow, 1993; Pascual, 1993; Truscott and Brindley, 1994). The problems of spatial and spatiotemporal pattern formation in plankton include regular and irregular oscillations, propagating fronts, spiral waves, pulses and stationary spiral patterns. Some significant contributions are dynamical stabilization of unstable equilibria (Petrovskii and Malchow, 2000; Malchow and Petrovskii, 2002) and chaotic oscillations behind propagating diffusive fronts in prey-predator models with finite or slightly inhomogeneous distributions (Sherratt et al., 1995, 1997; Petrovskii and Malchow, 2001).

Conceptual prey-predator models have often and successfully been used to model phytoplankton-zooplankton interactions and to elucidate mechanisms of spatiotemporal pattern formation like patchiness and blooming (Segel and Jackson, 1972; Steele and Hunderson, 1981; Pascual, 1993; Malchow, 1993). The density of plankton population changes not only in time but also in space. The highly inhomogeneous spatial distribution of plankton in the natural aquatic system called ``Plankton patchiness" has been observed in many field observations (Fasham, 1978; Steele, 1978; Mackas and Boyd, 1979; Greene et al., 1992; Abbott, 1993). Patchiness is affected by many factors like temperature, nutrients and turbulence, which depend on the spatial scale. Generally the growth, competition, grazing and propagation of plankton population can be modeled by partial differential equations of reaction-diffusion type. The fundamental importance of spatial and spatiotemporal pattern formation in biology is self-evident. A wide variety of spatial and temporal patterns inhibiting dispersal are present in many real ecological systems. The spatiotemporal self-organization in prey-predator communities modelled by reaction-diffusion equations have always been an area of interest for researchers. Turing spatial patterns have been observed in computer simulations of interaction-diffusion system by many authors (Malchow, 1996, 2000; Xiao et al., 2006; Brentnall et al., 2003; Grieco et al., 2005; Medvinsky et al., 2001; 2005; Chen and Wang, 2008; Liu et al., 2008). Upadhyay et al. (2009, 2010) investigated the wave phenomena and nonlinear non-equilibrium pattern formation in a spatial plankton-fish system with Holling type II and IV functional responses.

In this paper, we have tried to find out the analytical solution to plankton pattern formation which is in fact necessity for understanding the complex dynamics of a spatial plankton system with Holling type III functional response. And the analytical finding is supported by extensive numerical simulation.


\vspace*{0.5cm}
\setcounter{equation}{0}
\section{ The model system}

We consider a reaction-diffusion model for phytoplankton-zooplankton-fish system where at any point $(x, y)$ and time $t$, the phytoplankton $P(x, y, t)$ and zooplankton  $H(x, y, t)$ populations. The phytoplankton population $P(x, y, t)$ is predated by the zooplankton population $H(x, y, t)$ which is predated by fish. The per capita predation rate is described by Holling type III functional response. The system incorporating effects of fish predation satisfy the following:
\begin{equation}\label{21}
\begin{array}{l}
\frac{\partial P}{\partial t}=rP-B_1P^2-\frac{B_2P^2H}{P^2+D^2}+d_1\nabla^2P,\\[6pt]
\frac{\partial H}{\partial t}=C_1H-C_2\frac{H^2}{P}-\frac{FH^2}{H^2+D_1^2}+d_2\nabla^2H
\end{array}
\end{equation}
with the non-zero initial conditions
\begin{equation}\label{22}
\begin{array}{l}
P(x,y,0)>0,\,\, H(x,y,0)>0,\qquad (x,y)\in\Omega=[0, R]\times[0, R],
\end{array}
\end{equation}
and the zero-flux boundary conditions
\begin{equation}\label{23}
\frac{\partial P}{\partial n}=\frac{\partial H}{\partial n}=0,\,
(x,y)\in\partial\Omega\quad \text{for all}\quad t.
\end{equation}
where, $n$ is the outward unit normal vector of the
boundary $\partial \Omega$ which we will assume is smooth.
And $\nabla^2$ is the Laplacian operator, in two-dimensional space, $\nabla^2=\frac{\partial^2}{\partial x^2}+\frac{\partial^2}{\partial y^2}$, while in one-dimensional case, $\nabla^2=\frac{\partial^2}{\partial x^2}$.

The parameters $r, B_1, B_2, D, C_1, C_2, D_1, d_1, d_2$ in model~\eqref{21} are positive constants. We explain the meaning of each variable and constant. $r$ is the prey's intrinsic growth rate in the absence of predation, $B_1$ is the intensity of competition among individuals of  phytoplankton, $B_2$ is the rate at which phytoplankton is grazed by zooplankton and it follows Holling type--III functional response, $C_1$ is the predator's intrinsic rate of population growth, $C_2$  indicates the number of prey necessary to support and replace each individual predator. The rate equation for the zooplankton population is the logistic growth with carrying capacity proportional to phytoplankton density, $P/C_2$.  $D, D_1$ is the half-saturation constants for phytoplankton and zooplankton density respectively, $F$  is the maximum value of the total loss of zooplankton due to fish predation, which also follows Holling type-III functional response.  $d_1, d_2$ diffusion coefficients of phytoplankton and zooplankton respectively. The units of the parameters are as follows. Time $t$ and length $x, y\in[0, R]$  are measured in days $[d]$ and meters $[m]$ respectively. $r, P, H, D$ and $D_1$ are usually measured in $mg$ of dry weight per litre [$mg\cdot dw/l$]; the dimension of $B_2$, and $C_2$  is [$d^{-1}$],  are measured in [$(mg\cdot dw/l)^{-1}d^{-1}$] and [$d^{-1}$], respectively. The diffusion coefficients $d_1$  and $d_2$ are measured in [$m^{2}d^{-1}$].  $F$ is measured in [$(mg\cdot dw/l)d^{-1}$]. The significance of the terms on the right hand side of Eq.(1) is explained as follows: The first term represents the density-dependent growth of prey in the absence predators. The third term on the right of the prey equation, $B_2P^2/(P^2+D^2)$, represents the action of the zooplankton. We assume that the predation rate follows a sigmoid (Type III) functional response. This assumption is suitable for plankton community where spatial mixing occur due to turbulence (Okobo, 1980). The dimensions and other values of the parameters are chosen from literatures (Malchow, 1996; Medvinsky et al., 2001, 2002 Murray, 1989), which are well established for a long time to explain the phytoplankton-zooplankton dynamics. Notably, the system (1) is a modified (Holling III instead of II) and extended (fish predation) version of a model proposed by May (1973).

Holling type III functional response have often been used to demonstrate cyclic
collapses which form is an obvious choice for representing the behavior of predator
hunting (Real, 1977; Ludwig et al., 1978). This response function is sigmoid, rising
slowly when prey are rare, accelerating when they become more abundant, and finally
reaching a saturated upper limit. Type III functional response also levels off at some prey
density. Keeping the above mentioned properties in mind, we have considered the
zooplankton grazing rate on phytoplankton and the zooplankton predation by fish follows
a sigmoidal functional response of Holling type III.

\vspace*{0.5cm}
\setcounter{equation}{0}
\section{Stability analysis of the non-spatial model system}

In this section, we restrict ourselves to the stability analysis of the model system in the
absence of diffusion in which only the interaction part of the model system are taken into account. We find the non-negative equilibrium states of the model system and discuss
their stability properties with respect to variation of several parameters.

\subsection{Local stability analysis}

We analyze model system~\eqref{21} without diffusion. In such case, the model system reduces to
\begin{equation}\label{31}
\begin{array}{l}
\frac{\partial P}{\partial t}=rP-B_1P^2-\frac{B_2P^2H}{P^2+D^2},\\[6pt]
\frac{\partial H}{\partial t}=C_1H-C_2\frac{H^2}{P}-\frac{FH^2}{H^2+D_1^2}
\end{array}
\end{equation}
with $P(x, 0)>0, H(x, 0)>0$.

The stationary dynamics of system~\eqref{31} can be analyzed from $dP/dt = 0, dH/dt=0$.
Then, we have
\begin{equation}\label{32}
\begin{array}{l}
rP-B_1P^2-\frac{B_2P^2H}{P^2+D^2}=0,\\[6pt]
C_1H-C_2\frac{H^2}{P}-\frac{FH^2}{H^2+D_1^2}=0.
\end{array}
\end{equation}
An application of linear stability analysis determines the stability of the two stationary
points, namely $E_1(\frac{r}{B_1}, 0)$ and $E^*(P^*, H^*)$.

Notably, $E^*(P^*, H^*)$ presents the kind of nontrivial coexistence equilibria, not only
mean one equilibrium. In fact, it may be noted that $P^*$ and $H^*$ satisfies the following
equations:
$$H^*-\frac{(r-B_2P^*)(P^{*2}+D^2)}{B_2P^*}=0,\, \frac{FH^*}{H^{*2}+D_1^2}+C_2\frac{H^*}{P^*}-C_1=0.$$
After solving $P^*$ from the second equation above and substituting it into the first
equation, we can obtain a polynomial equation about $H^*$ with degree nine; therefore the
equation at least has one real root. That is to say, system~\eqref{31} may be having one or more positive equilibria, namely $E^*(P^*, H^*)$.

From Eqs.~\eqref{31}, it is easy to show that the equilibrium point $E_1$ is a saddle point with stable manifold in $P-$direction and unstable manifold in $H-$direction. The stationary point
$E^*(P^*, H^*)$ is stable or unstable depend on the coexistence of phytoplankton and
zooplankton. In the following theorem we are able to find necessary and sufficient
conditions for the positive equilibrium $E^*(P^*, H^*)$ to be locally asymptotically stable.

The Jacobian matrix of model~\eqref{31} at $E^*(P^*, H^*)$ is
$$J=\left(\begin{array}{cc}P^*\left(-B_1+B_2H^*f_1(P^*, D)\right) &
-\frac{B_2P^{*2}}{P^{*2}+D^2} \\
\frac{C_2H^{*2}}{P^{*2}}&H^*\left(-\frac{C_2}{P^*}+Ff_2(H^*, D_1)\right)\end{array}
\right),$$
where,
$$f_1(P^*, D)=\frac{P^{*2}-D^2}{(P^{*2}+D^2)^2},\quad f_2(H^*, D)=\frac{H^{*2}-D_1^2}{(H^{*2}+D_1^2)^2}.$$

Following the Routh-Hurwitz criteria, we get,
\begin{equation}\label{33}
\begin{array}{l}
A=P^*(B_1-B_2H^*f_1(P^*, D))+H^*(\frac{C_2}{P^*}-Ff_2(H^*, D_1)),\\[6pt]
B=P^*H^*(B_1-B_2H^*f_1(P^*, D))\left(\frac{C_2}{P^*}-Ff_2(H^*, D_1)\right)+\frac{B_2C_2H^{*2}}{(P^{*2}+D^2)^2}.
\end{array}
\end{equation}

Summarizing the above discussions, we can obtain the following theorem.

\vspace*{0.25cm}
\begin{theorem}
The positive equilibrium $E^*(P^*, H^*)$ is locally asymptotically stable in the
$PH-$plane if and only if the following inequalities hold:
$$A > 0 \quad \text{and}\quad  B > 0.$$
\end{theorem}

\vspace*{0.25cm}

\begin{remark}
Let the following inequalities hold
\begin{equation}\label{34}
\begin{array}{l}
B_1>B_2H^*f_1(P^*, D),\quad
C_2>FP^*f_2(H^*, D_1).
\end{array}
\end{equation}
Then $A>0$ and $B>0$. It shows that if inequalities in Eq.~\eqref{34} hold, then $E^*(P^*, H^*)$ is locally asymptotically stable in the $PH-$plane.
\end{remark}

Fig.1 shows the dynamics of system~\eqref{31} that for increasing fish predation rates at $F=0$
(limit cycle---no fish), $F=0.02$, 0.03, 0.04 (limit cycle) and $F=0.044$ (phytoplankton
dominance). We observe the limit cycle behavior for the fish predation rate $F$ in the
range $0\leq F\leq 0.043$, and for higher values phytoplankton dominance is reported for a
non-spatial system~\eqref{31}.

\begin{figure}[htbp]
\centerline{\includegraphics[scale=0.8]{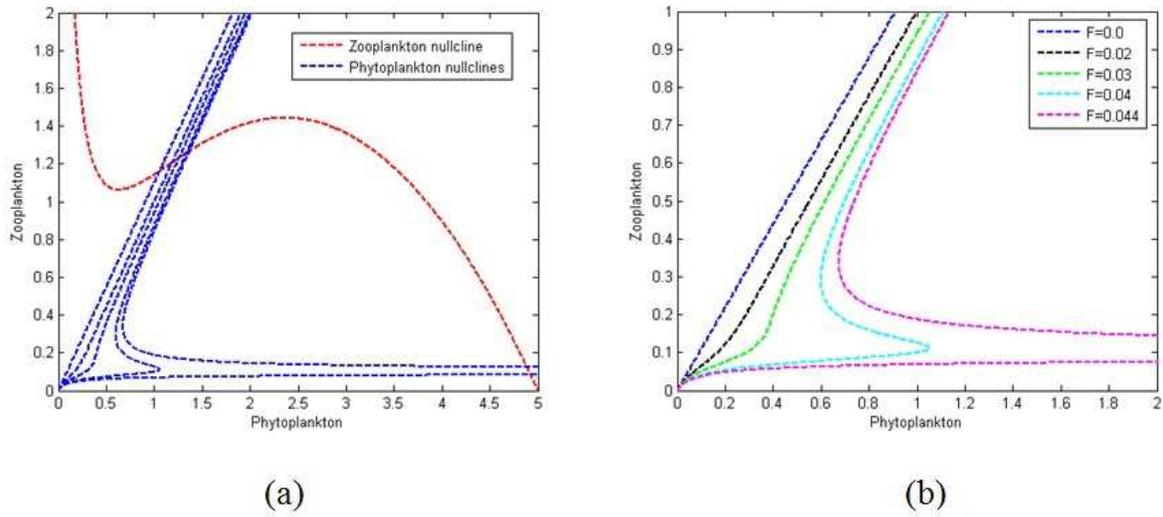}}
\caption{(a) predator-prey zero-isoclines (b) prey zero-isoclines for different value of $F$ with parameters value $r=1$, $B_1=0.2$, $B_2=0.91$, $C_1=0.22, C_2=0.2$, $D^2=0.3$, $D_1=0.1$.}
 \label{figure1}
\end{figure}


\subsection{Global stability analysis}

In order to study the global behavior of the positive equilibrium $E^*(P^*, H^*)$, we need
the following lemma which establishes a region for attraction for model system~\eqref{31}.

\vspace*{0.25cm}

\noindent{\bf Lemma 1.}
The set $\mathcal{R}=\{(P, H): 0\leq P\leq \frac{r}{B_1}, 0\leq H\leq\frac{rC_1}{C_2B_1}\}$ is a region of attraction for all solutions initiating in the interior of the positive quadrant $\mathbb{R}$.

\vspace*{0.25cm}

\begin{theorem}
If
\begin{equation}\label{35}
P^*(r-B_1P^*)^2<4B_1rD^2,\quad rC_1H^*<D_1^2C_2B_1
\end{equation}
hold, then $E^*$ is globally asymptotically stable with respect to all solutions in the interior of the positive quadrant $\mathbb{R}$.
\end{theorem}

\noindent {\bf Proof.}  Let us choose a Lyapunov function
\begin{equation}\label{Lyap}V(P, H)=\int_{P^*}^P\frac{\xi-P^*}{\xi\varphi(\xi)}d\xi+w\int_{H^*}^H\frac{\eta-H^*}{\eta}d\eta,
\end{equation}
where $\varphi(P)=\frac{B_2P^2}{P^2+D^2}$.

Differentiating both sides with respect to $t$, we get
$$\frac{dV}{dt}(P, H)=\frac{P-P^*}{P\varphi(P)}\frac{dP}{dt}+w\frac{H-H^*}{H}\frac{dH}{dt},$$
Substituting the expressions of
$\frac{dP}{dt}$ and $\frac{dH}{dt}$
from Eq.~\eqref{31} and putting
$w=\frac{P^*}{C_2H^*}$, we obtained:
$$\begin{array}{l}
\frac{dV}{dt}=\frac{P-P^*}{P}\Bigl(\frac{rP-B_1P^2}{\varphi(P)}-H^*\Bigr)-(H-H^*)^2\frac{P^*}{H^*P}
-w(H-H^*)^2\frac{F(D_1^2-H^*H)}{(H^2+D_1^2)(H^{*2}+D_1^2)}\\[8pt]
\quad\,\, =\frac{(P-P^*)^2}{BP^2P^*}\Bigl(rP^*P-rD^2-B_1PP^*(P^*+P)\Bigr)-(H-H^*)^2\frac{P^*}{H*P}\\[8pt]
\qquad\,\,\,\, -w(H-H^*)^2\frac{F(D_1^2-H^*H)}{(H^2+D_1^2)(H^{*2}+D_1^2)}
\end{array}$$
then $\frac{dV}{dt}<0$
for all $P, H\in\mathcal{R}$ if the condition~\eqref{35} hold.

\vspace*{0.5cm}
\setcounter{equation}{0}
\section{Stability analysis of the spatial model system}

In this section, we study the effect of diffusion on the model system about the interior
equilibrium point. Complex marine ecosystems exhibit patterns that are bound to each
other yet observed over different spatial and time scales (Grimm et al., 2005). Turing instability can occur for the model system because the equation for predator is nonlinear
with respect to predator population, $H$ and unequal values of diffusive constants.
To study the effect of diffusion on the system~\eqref{31}, we derive conditions for stability
analysis in one and two dimensional cases.

\subsection{One dimensional case}

The model system~\eqref{21} in the presence of one-dimensional diffusion has the following
form:
\begin{equation}\label{41}
\begin{array}{l}
\frac{\partial P}{\partial t}=rP-B_1P^2-\frac{B_2P^2H}{P^2+D^2}+d_1\frac{\partial^2P}{\partial x^2},\\[6pt]
\frac{\partial H}{\partial t}=C_1H-C_2\frac{H^2}{P}-\frac{FH^2}{H^2+D_1^2}+d_2\frac{\partial^2H}{\partial x^2}.
\end{array}
\end{equation}

To study the effect of diffusion on the model system, we have considered the
linearized form of system about $E^*(P^*, H^*)$ as follows:
\begin{equation}\label{42}
\begin{array}{l}
\frac{\partial U}{\partial t}=b_{11}U+b_{12}V+d_1\frac{\partial^2U}{\partial x^2},\\[8pt]
\frac{\partial V}{\partial t}=b_{21}U+b_{22}V+d_2\frac{\partial^2V}{\partial x^2},
\end{array}
\end{equation}
where $P=P^*+U, H=H^*+V$ and
$$\begin{array}{l}
b_{11}=-P^*(B_1-B_2H^*f_1(P^*, D)),\quad b_{12}=-B_2P^*/(P^{*2}+D^2),\\[6pt]
b_{21}=C_2H^{*2}/P^{*2},\qquad\qquad\qquad\quad\,\,\,\,\, b_{22}=H^*(Ff_2(H^*, D_1)-C_2/P^*).
\end{array}$$

It may be noted that $(U, V)$ are small perturbations of $(P, H)$ about the equilibrium point
$E^*(P^*, H^*)$.

In this case, we look for eigenfunctions of the form
$$\sum_{n=0}^{\infty}{a_n\choose b_n}\exp{(\lambda t+ikx)},$$
and thus solutions of system~\eqref{42} of the form
$${U\choose V}=\sum_{n=0}^{\infty}{a_n\choose b_n}\exp{(\lambda t+ikx)},$$
where $\lambda$ and $k$ are the frequency and wave-number respectively. The characteristic
equation of the linearized system~\eqref{42} is given by
\begin{equation}\label{43}
\lambda^2+\rho_1\lambda+\rho_2=0,
\end{equation}
where
\begin{equation}\label{44}
\rho_1=A+(d_1+d_2)k^2,
\end{equation}
\begin{equation}\label{45}
\rho_2=B+d_1d_2k^4+\Bigl[d_2P^*(B_1-B_2H^*f_1(P^*, D))+d_1H^*(C_2/P^*-Ff_2(H^*, D_1))\Bigr]k_2,
\end{equation}
where $A$ and $B$ are defined in Eq.~\eqref{33}. From Eqs.~\eqref{43}---\eqref{45}, and using Routh-Hurwitz criteria, we can know that the positive equilibrium
$E^*$ is locally asymptotically stable in the presence of diffusion if and only if
\begin{equation}\label{46}
\rho_1>0\quad \text{and}\quad \rho_2>0.
\end{equation}

Summarizing the above discussions, we can get the following theorem immediately.

\vspace*{0.25cm}
\begin{theorem}

(i) If the inequalities in Eq.~\eqref{34} are satisfied, then the positive equilibrium point $E^*$ is locally asymptotically stable in the presence as well as absence of diffusion.

(ii) Suppose that inequalities in Eq.~\eqref{34} are not satisfied, i.e., either $A$ or $B$ is negative or both $A$ and $B$ are negative. Then for strictly positive wave-number $k > 0$, i.e., spatially inhomogeneous perturbations, by increasing $d_1$ and
$d_2$ to sufficiently large values, $\rho_1$
and $\rho_2$ can be made positive and hence $E^*$ can be made locally asymptotically stable.
\end{theorem}

Diffusive instability sets in when at least one of the conditions in Eq.~\eqref{46} is
violated subject to the conditions in Theorem 1. But it is evident that the first condition
$\rho_1 > 0$ is not violated when the condition $A > 0$ is met. Hence only the violation of
condition $\rho_2>0$ gives rise to diffusion instability. Hence the condition for diffusive
instability is given by
\begin{equation}\label{47}
H(k^2)=d_1d_2k^4+\Bigl[d_2P^*(B_1-B_2H^*f_1(P^*, D))+d_1H^*(C_2/P^*-Ff_2(H^*, D_1))\Bigr]k_2+B<0.
\end{equation}
$H$ is quadratic in $k^2$ and the graph of $H(k^2)=0$ is a parabola. The minimum of $H(k^2)$
occurs at $k^2=k_m^2$ where
\begin{equation}\label{48}
k_m^2=\frac{1}{2d_1d_2}\Bigl[d_2P^*(B_2H^*f_1(P^*, D)-B_1)+d_1H^*(Ff_2(H^*, D_1)-C_2/P^*)\Bigr]>0.
\end{equation}
Consequently the condition for diffusive instability is $H(k_m^2)$. Therefore
\begin{equation}\label{49}
\frac{1}{4d_1d_2}\Bigl[d_2P^*(B_2H^*f_1(P^*, D)-B_1)+d_1H^*(Ff_2(H^*, D_1)-C_2/P^*)\Bigr]>B.
\end{equation}

\vspace*{0.25cm}
\begin{theorem}
The criterion for diffusive instability for the model system is obtained by combining the
result of Theorem 1, \eqref{48} and \eqref{49} and leading to the following condition:
\begin{equation}\label{410}
d_2P^*(B_2H^*f_1(P^*, D)-B_1)+d_1H^*(Ff_2(H^*, D_1)-C_2/P^*>2(Bd_1d_2)^{\frac{1}{2}}>0.
\end{equation}
\end{theorem}

In the following theorem, we are able to show the global stability behaviour of the
positive equilibrium in the presence of diffusion.

\vspace*{0.25cm}
\begin{theorem}
(i) If the equilibrium $E^*$ of system~\eqref{31} is globally asymptotically stable, the
corresponding uniform steady state of system~\eqref{42} is also globally asymptotically stable.

(ii) If the equilibrium $E^*$ of system~\eqref{31} is unstable even then the corresponding uniform steady state of system~\eqref{42} can be made globally asymptotically stable by increasing the diffusion coefficient $d_1$ and $d_2$ to a sufficiently large value for strictly positive wave-number $k > 0$.
\end{theorem}

\noindent {\bf Proof.} For the sake of simplicity, let $P(x, t)=P, H(x, t)=H$. For $x\in[0, R]$ and $t\in[0, \infty]$, let us define a functional:
$$V_1=\int_0^RV(P, H)dx$$
where $V(P, H)$ is defined as Eq.~\eqref{Lyap}.

Differentiating $V_1$ with respect to time $t$ along the solutions of system~\eqref{42}, we get
$$\frac{dV_1}{dt}=\int_0^R\Bigl(\frac{\partial V}{\partial P}\frac{\partial P}{\partial t}+\frac{\partial V}{\partial H}\frac{\partial H}{\partial t}\Bigr)dx
=\int_0^R\frac{dV}{dt}dx+\int_0^R
\Bigl(d_1\frac{\partial V}{\partial P}\frac{\partial^2 P}{\partial x^2}+d_2\frac{\partial V}{\partial H}\frac{\partial^2 H}{\partial x^2}\Bigr)dx.$$

Using the boundary condition~\eqref{23}, we obtain
\begin{equation}\label{411}
\frac{dV_1}{dt}=\int_0^R\frac{dV}{dt}dx-d_1\int_0^R
\frac{P^*(P^2+3D^2)-2D^2P}{B_2P^4}\Bigl(\frac{\partial P}{\partial x}\Bigr)^2dx-d_2\int_0^R\frac{wH^*}{H^2}\Bigl(\frac{\partial H}{\partial x}\Bigr)^2dx.
\end{equation}
From Eq.~\eqref{411}, we note that if $\frac{dV}{dt}<0$ then $\frac{dV_1}{dt}$ in the interior of the positive quadrant of the $PH-$plane, and hence the first part of the theorem follows. We also note that if $\frac{dV}{dt}<0$ then $\frac{dV_1}{dt}$
can be made negative by increasing $d_1$ and $d_2$ a sufficiently
large value, and hence the second part of the theorem follows.

\subsection{Two dimensional case}

In two-dimensional case, the model system~\eqref{21} can be written as

\begin{equation}\label{412}
\begin{array}{l}
\frac{\partial P}{\partial t}=rP-B_1P^2-\frac{B_2P^2H}{P^2+D^2}+d_1\Bigl(\frac{\partial^2P}{\partial x^2}+\frac{\partial^2P}{\partial y^2}\Bigr),\\[6pt]
\frac{\partial H}{\partial t}=C_1H-C_2\frac{H^2}{P}-\frac{FH^2}{H^2+D_1^2}+d_2\Bigl(\frac{\partial^2H}{\partial x^2}+\frac{\partial^2H}{\partial y^2}\Bigr).
\end{array}
\end{equation}

In this section, we show that the result of theorem 5 remain valid for two dimensional
case. To prove this result we consider the following functional (Dubey and Hussain,
2000):
$$V_2=\iint\limits_{\Omega}V(P, H)dA$$
where $V(P, H)$ is defined as Eq.~\eqref{Lyap}.

Differentiating $V_2(t)$ with respect to time $t$ along the solutions of system~\eqref{412}, we obtain
\begin{equation}\label{413}
\frac{dV_2}{dt}=I_1+I_2,
\end{equation}
where
$$I_1=\iint\limits_{\Omega}\frac{dV}{dt}dA,\qquad I_2=\iint\limits_{\Omega}\Bigl(d_1\frac{\partial V}{\partial P}\nabla^2P+d_2\frac{\partial V}{\partial H}\nabla^2H\Big)dA.$$
Using Green's first identity in the plane
$$\iint\limits_{\Omega}F\nabla^2GdA=\int\limits_{\partial\Omega}F\frac{\partial G}{\partial
n}ds-\iint\limits_{\Omega}(\nabla F\cdot \nabla G)dA.$$
and under an analysis similar to Dubey and Hussain(2000), one can show that
$$d_1\iint\limits_{\Omega}\left(\frac{\partial V}{\partial P}\nabla^2 P\right)dA
=-d_1\iint\limits_{\Omega}\frac{\partial^2V}{\partial P^2}\left[(\frac{\partial P}{\partial x})^2+(\frac{\partial P}{\partial y})^2\right]dA\leq 0,$$
$$d_2\iint\limits_{\Omega}\left(\frac{\partial V}{\partial H}\nabla^2 H\right)dA
=-d_2\iint\limits_{\Omega}\frac{\partial^2V}{\partial H^2}\left[(\frac{\partial H}{\partial x})^2+(\frac{\partial H}{\partial y})^2\right]dA\leq 0.$$
This shows that $I_2\leq 0$. From the above analysis we note that if $I_1<0$, then
$\frac{dV_2}{dt}$.
This implies that if in the absence of diffusion $E^*$ is globally asymptotically stable, then
in the presence of diffusion $E^*$ will remain globally asymptotically stable.

We also note that if $\frac{dV}{dt}>0$, then $I_1>0$. In such a case diffusion, $E^*$  will be unstable in the absence of diffusion. Even in this case by increasing $d_1$ and $d_2$ to a sufficiently large value $\frac{dV_2}{dt}$ can be made negative. This shows that if in the absence of diffusion $E^*$ is unstable, then in the presence of diffusion  $E^*$ can be made stable by increasing diffusion coefficients to sufficiently large value.

\vspace*{0.5cm}
\setcounter{equation}{0}
\section{Numerical simulations}

In this section, we perform numerical simulations to illustrate the results obtained in
previous sections. The dynamics of the model system~\eqref{21} is studied with the help of
numerical simulation, both in one and two dimensions, to investigate the spatiotemporal
dynamics of the model system~\eqref{21}. For the one-dimensional case, the plots (space vs.
population densities) are obtained to study the spatial dynamics of the model system. The
temporal dynamics is studied by observing the effect of time on space vs. density plot of
prey populations. For the two-dimensional case, the spatial snapshots of prey densities
are obtained at different time levels for different values of $F$ and we have tried to study
the spatiotemporal dynamics of the spatial model system. The rate of fish predation, $F$
acting as forcing term or distributed control parameter in the model system~\eqref{21} which is
strictly nonnegative. Mathematically, we assume that $F$ can be manipulated at every point
in space and time. However, from practical point of view, we only have direct control of
the net density of fish population at any instant. The parameters determining the shape of
the type-III functional response of fish to zooplankton density are set more or less
arbitrarily. They will in fact very much depend on the fish species, and also change
significantly within a species with the size of the individuals.

\subsection{Spatiotemporal dynamics in One-dimensional system}

Now we start our insight into the spatiotemporal dynamics of the model system~\eqref{21} in one-dimension with the following set of parameter values. The spatiotemporal dynamics of
the system depends to a large extent on the choice of initial conditions. In a real aquatic
ecosystem, the details of the initial spatial distribution of the species can be caused by
spatially homogenous initial conditions. However, in this case, the distribution of species
would stay homogenous for any time, and no spatial pattern can emerge. To get a
nontrivial spatiotemporal dynamics, we have perturbed the homogenous distribution. We
start with a hypothetical ``constant-gradient" distribution (Malchow et al., 2008):
\begin{equation}\label{51}
P(x,0)=P^*,\quad H(x,0)=H^*+\varepsilon x+\delta,
\end{equation}
where $\varepsilon$ and $\delta$ are parameters. $(P^*, H^*)=(1.8789, 1.3984)$ is the non-trivial equilibrium point of system~\eqref{31}, with fixed parameter set
$$r=1, B_1=0.2, B_2=0.91, C_1=0.22, C_2=0.2, D^2=0.3, D_1=0.1, F=0.1.$$

It appears that the type of the system dynamics depends on $\varepsilon$ and $\delta$. In Figs. 2 and 3, we show the population distribution over space at time $t = 2000$ (Fig.2) and $t=1000, 2000$(Fig.3), respectively. In Fig. 4, we illustrate the pattern formation about the above two cases, the system size is 3000, iteration numbers from 17800 to 19800, i.e., time $t$ from 1780 to 1980. In Fig.2a and Fig.4a, $\varepsilon=-1.5\times 10^{-6}$ and $\delta=10^{-6}$, the spatial distributions gradually vary in time, and the local temporal behavior of the dynamical variables $P$ and $H$ is strictly periodic following limit cycle of the non-spatial system, which is perhaps what is intuitively expected from system~\eqref{41}, there emerge a periodic wave pattern (Fig.4a, for $P$). However, in Fig.2b and Fig.4b, for a slightly different set of $\varepsilon$ and $\delta$, i.e.,
$\varepsilon=-1.5\times 10^{-3}$ and $\delta=10^{-6}$, the dynamics of the system undergoes principal changes, and there emerge an oscillations wave pattern with a ``flow" pattern around $x=1500$ (c.f., Fig.4b). In Fig.3 and Fig.4c, $\varepsilon=-1.5\times 10^{-2}$ and $\delta=10^{-3}$, more different with the two cases above,
the initial conditions~\eqref{51} lead to the formation of strongly irregular ``jagged" dynamics pattern inside a sub-domain of the system. There is a growing region of periodic
travelling waves ($x = 2000$ to $x =2500$) followed by oscillations with a large spatial
wavelength. This is a typical scenario which precedes either the formation of
spatiotemporal irregular oscillations or spatially homogeneous oscillations (c.f., Fig.4c).

\begin{figure}[htbp]
\centerline{\includegraphics[scale=0.6]{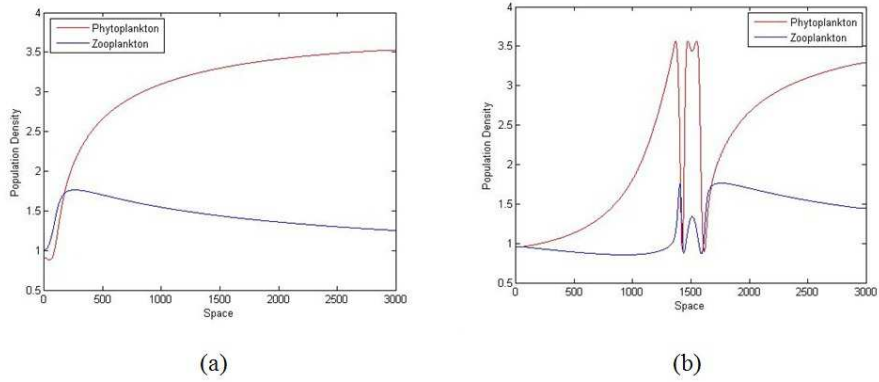}}
\caption{Population distributions over space at $t = 2000$ obtained for the parameter values  $r=1, B_1=0.2, B_2=0.91, C_1=0.22, C_2=0.2, D^2=0.3, D_1=0.1, F=0.1$ and the initial conditions ~\eqref{51} with (a) $\varepsilon=-1.5\times 10^{-6}$ and $\delta=10^{-6}$;  (b) $\varepsilon=-1.5\times 10^{-3}$ and $\delta=10^{-6}$.}
 \label{figure2}
\end{figure}
\begin{figure}[htbp]
\centerline{\includegraphics[scale=0.6]{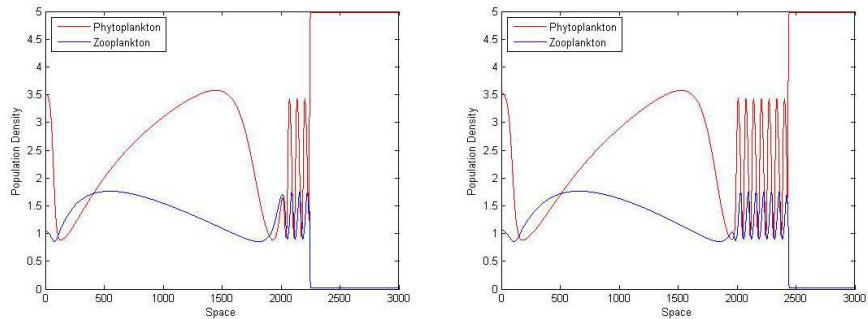}}
\caption{Population distributions over space at (a) $t=1000$ (b) $t=2000$  obtained for the parameter values and the initial conditions~\eqref{51} with $\varepsilon=-1.5\times 10^{-2}$ and $\delta=10^{-3}$. The values of the other parameters are the same as given in Fig. 2.}
 \label{figure3}
\end{figure}
\begin{figure}[htbp]
\centerline{\includegraphics[scale=0.8]{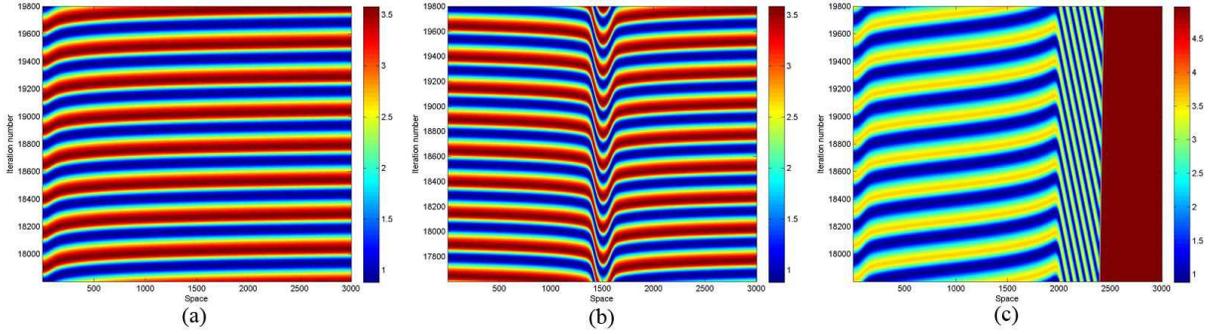}}
\caption{Spatiotemporal dynamical of system~\eqref{41} with the initial conditions~\eqref{51}. Parameters: (a) $\varepsilon=-1.5\times 10^{-6}$, $\delta=10^{-6}$;  (b) $\varepsilon=-1.5\times 10^{-3}$, $\delta=10^{-6}$; (c) $\varepsilon=-1.5\times 10^{-2}$, $\delta=10^{-3}$.}
 \label{figure4}
\end{figure}

\subsection{Spatiotemporal dynamics in Two-dimensional system}

In this subsection, we show the spatiotemporal dynamics in two-dimensional case of
system~\eqref{412}.

\subsubsection{Turing pattern}

First, we show Turing pattern of system~\eqref{412} obtained by performing numerical
simulations with initial and boundary conditions given in Eqs.~\eqref{22}--\eqref{23} with the
parameter values given below in Eq. (27) with system size $200\times 200$ for different values
of $F$ and time $t$. The values of the other parameters are fixed as
\begin{equation}\label{52}
\begin{array}{l}
r=1, B_1=0.2, B_2=0.91, C_1=0.22, C_2=0.2, D^2=0.3, D_1=0.1, \\[6pt]
d_1=0.005, d_2=1,
\Delta x=\Delta y=1/3, \Delta t=1/40.
\end{array}
\end{equation}

Initially, the entire system is placed in the stationary state
$(P^*, H^*)$, and the propagation velocity of the initial
perturbation is thus on the order of $5\times 10^{-4}$ space units
per time unit. And the system is then integrated for $10^5$ or $2\times 10^5$ time steps and some images saved. After the initial period
during which the perturbation spreads, either the system goes into a
time dependent state, or to an essentially steady state (time
independent). Here, we show the
distribution of phytoplankton $P$, for instance.

Fig.5 shows five typical Turing pattern of phytoplankton $P$ in system~\eqref{412} with
parameters set~\eqref{52}. From Fig. 5, one can see that values for the concentration P are
represented in a color scale varying from blue (minimum) to red (maximum), and on
increasing the control parameter $F$, the sequences spots $\rightarrow$ spot-stripe mixtures$\rightarrow$ stripes$\rightarrow$ hole-stripe mixtures holes pattern is observed. When $F > 0.077$, other parameters are
the same as above case, wave pattern emerges and are presented in Fig. 6.
\begin{figure}[htbp]
\centerline{\includegraphics[scale=0.8]{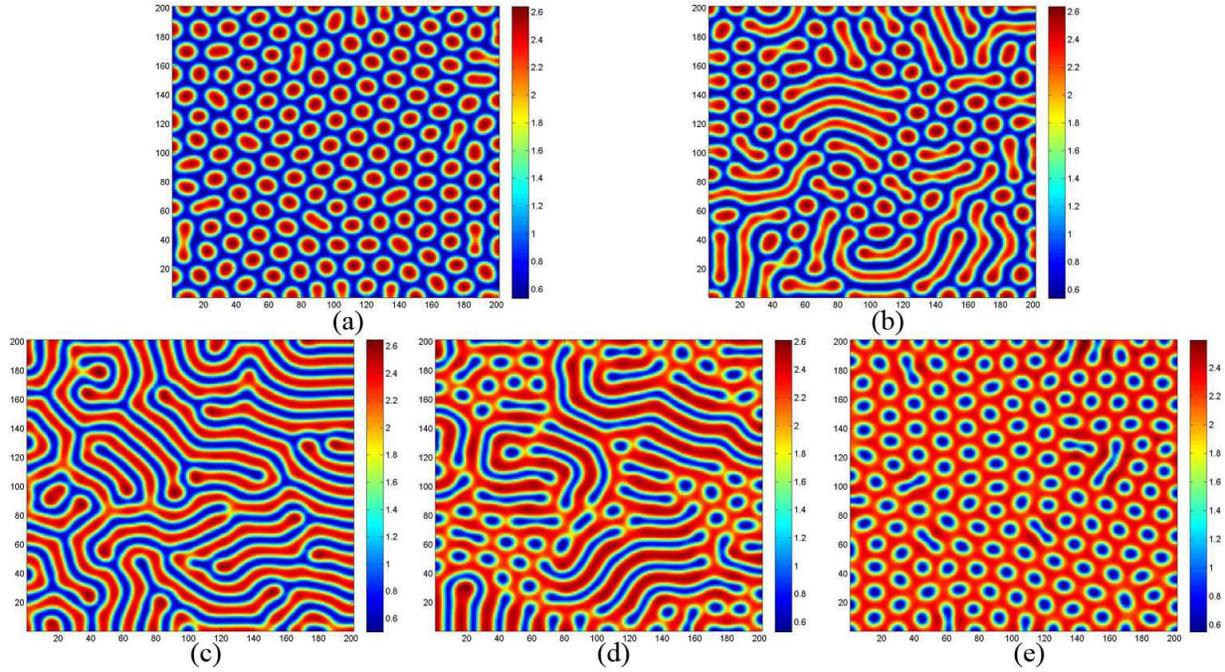}}
\caption{Typical Turing patterns of $P$ in the system~\eqref{412} with fixed parameters set ~\eqref{52}. (a) spots pattern, $F=0.0001$; (b) spot-stripe mixtures pattern, $F=0.01$; (c) stripes pattern, $F=0.035$; (d) hole-stripe mixtures, $F=0.065$; (e) holes pattern, $F=0.0765$. Iterations: Pattern-(a) and (e): $2\times 10^5$, others: $10^5$.}
 \label{figure5}
\end{figure}
\begin{figure}[htbp]
\centerline{\includegraphics[scale=0.52]{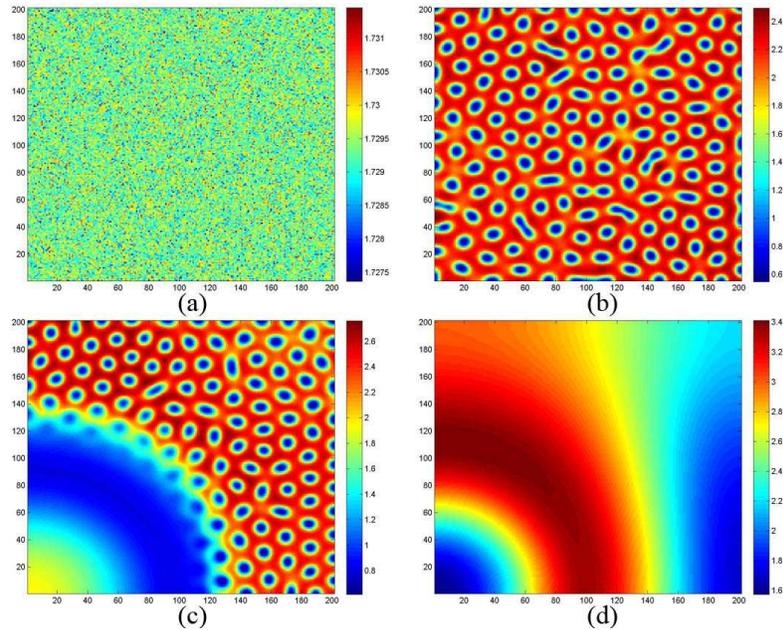}}
\caption{Wave pattern obtained with system~\eqref{412} with $F=0.085$. Iterations: (a) 0, (b) 10000, (c) 25000, (d) 40000.}
 \label{figure6}
\end{figure}

\subsubsection{Spiral wave pattern}

Thanks to the insightful work of Medvinsky et al. (2002), we have studied the spiral wave pattern for two different set of initial condition discussed in Eqs.~\eqref{53} and~\eqref{54} respectively. In these two cases, we employ $\Delta x=\Delta y=1, \Delta t=1/3$ and the system size is $900\times 300$. Other parameters set is fixed as~\eqref{52}.

The first case is:
\begin{equation}\label{53}
\begin{array}{l}
P(x,y,0)=P^*-\varepsilon_1(x-0.2y-225)(x-0.2y-675),\\[6pt] H(x,y,0)=H^*-\varepsilon_2(x-450)-\varepsilon_3(y-150).
\end{array}
\end{equation}
where $\varepsilon_1=2\times 10^{-7}, \varepsilon_2=3\times 10^{-5}, \varepsilon_3=1.2\times 10^{-4}$.

The initial conditions are deliberately chosen to be unsymmetrical in order to make any influence of the corners of the domain more visible. Snapshots of the spatial distribution arising from~\eqref{53} are shown in Fig. 7 for $t = 0, 2500, 5000, 10000$. Fig.7a shows that for the system~\eqref{412} with initial conditions~\eqref{53}, the formation of the irregular patchy structure can be preceded by the evolution of a regular spiral spatial pattern. Note that the appearance of the spirals is not induced by the initial conditions. The center of each spiral is situated in a critical point $(x_{cr}, y_{cr})$, where $U(x_{cr}, y_{cr}) = P^*$, $V(x_{cr}, y_{cr}) = H^*$. The distribution~\eqref{53} contains two such points; for other initial conditions, the number of spirals may be different. After the spirals form (Fig. 7b), they grow slightly for a certain time, their spatial structure becoming more distinct (Figs. 7c and 7d).

\begin{figure}[htbp]
\centerline{\includegraphics[scale=0.8]{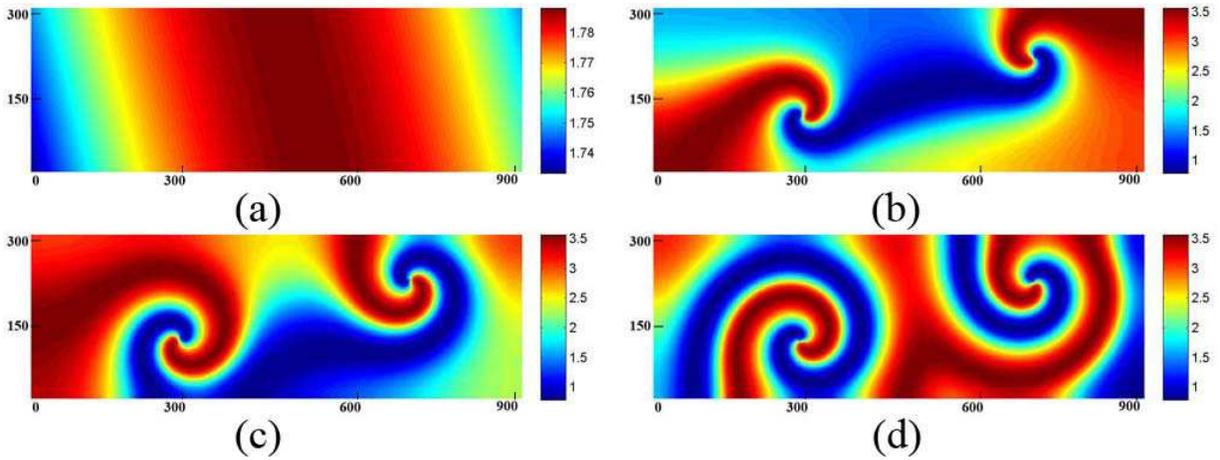}}
\caption{Spiral pattern of P in system~\eqref{412} with special initial condition~\eqref{53}  and $F=0.1$. Time: (a) 0, (b) 2500, (c) 5000, (d) 10000.}
 \label{figure7}
\end{figure}

In the second case, the initial conditions describe a phytoplankton (prey) patch placed into a domain with a constant-gradient zooplankton (predator) distribution:
\begin{equation}\label{54}
\begin{array}{l}
P(x,y,0)=P^*-\varepsilon_1(x-180)(x-720)-\varepsilon_2(y-90)(y-210),\\[6pt] H(x,y,0)=H^*-\varepsilon_3(x-450)-\varepsilon_4(y-135).
\end{array}
\end{equation}
where $\varepsilon_1=2\times 10^{-7}, \varepsilon_2=6\times 10^{-7}, \varepsilon_3=3\times 10^{-5}, \varepsilon_4=6\times 10^{-5}$.

Fig. 8 shows the snapshots of the spatial distribution arising from~\eqref{53} for $t = 0, 1000, 4000, 9000$. Although the dynamics of the system preceding the formation of patchy spatial structure is somewhat less regular than in the previous case, it follows a similar scenario. Again the spirals appear, with their centers located in the vicinity of critical points (Figs. 8c and 8d).

\begin{figure}[htbp]
\centerline{\includegraphics[scale=0.8]{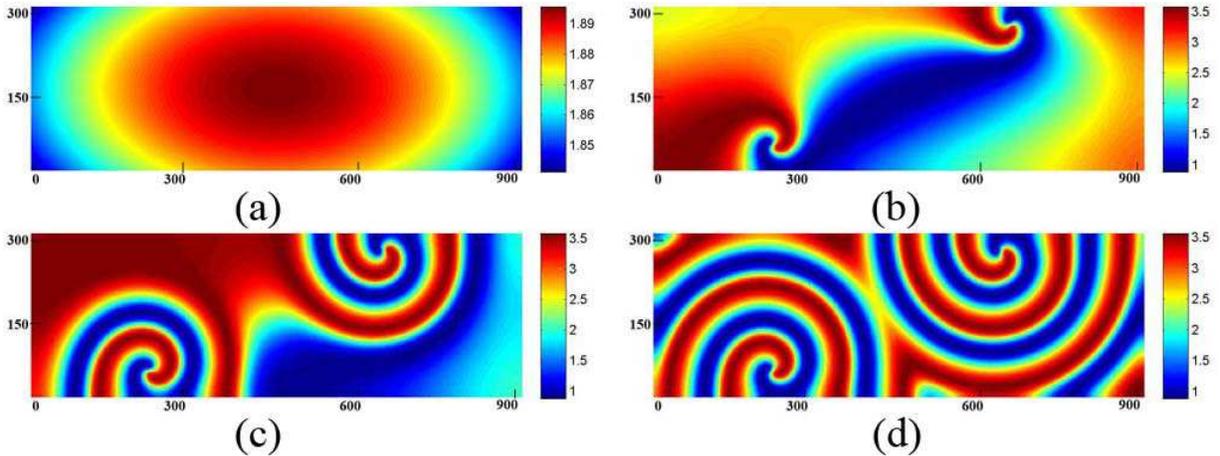}}
\caption{Spiral pattern of P in system~\eqref{412} with special initial condition~\eqref{54}  and $F=0.1$. Time: (a) 0, (b) 1000, (c) 4000, (d) 9000.}
 \label{figure7}
\end{figure}

\section{Discussions and Conclusions}

In this paper, we have made an attempt to provide a unified framework to understand the complex dynamical patterns observed in a spatial plankton model for phytoplankton-zooplankton-fish interaction with Holling type III functional response. Since most marine mammals and boreal fish species are considered to be generalist predators (including feeders, grazers, etc.), a sigmoidal (type III) functional response might be more appropriate (Magnusson and Palsson, 1991). This has been proposed as being likely for minke whales, harp seals, and cod in the Barents Sea where switching between prey species has been hypothesized (Nilssen et al., 2000; Schweder et al., 2000). Sigmoidal functional response arising due to heterogeneity of space is less addressed in the literature. It is well known that real vertical distribution of plankton is highly heterogeneous, especially during plankton blooms and this can seriously affect the behavior of plankton models (Poggiale et al. 2008). Morozov (2010) shows that emergence of Holling type III in plankton system is due to mechanisms different from those well known in the literature (e.g. food search learning by predator, existence of alternative food, refuse for prey). It has certainly been useful in both theoretical investigations and practical applications (Turchin, 2003). Movement of phytoplankton and zooplankton population with different velocities can give rise to spatial patterns (Malchow, 2000). We have studied the reaction-diffusion model in both one and two dimensions and investigated their stability conditions. By analyzing the model system~\eqref{21}, we observed that two points of equilibrium exist for the model system. The non-trivial equilibrium state  of prey-predator coexistence is locally as well as globally asymptotically stable under a fixed region of attraction when certain conditions are satisfied. It has been shown that the instability observed in the system is diffusion-driven under the condition~\eqref{410}. The expression for critical wave-number has been obtained. It has been observed that the solution of the model system converges faster to its equilibrium in case of diffusion in two-dimensional space as compared to the diffusion in one dimensional space.

In the numerical simulation, we adopt the rate of fish predation F in the system~\eqref{21} as the control parameter which is strictly nonnegative. In the two-dimensional case, on increasing the value of $F$, the sequences spots$\rightarrow$spot-stripe mixtures$\rightarrow$stripes$\rightarrow$ hole-stripe mixtures$\rightarrow$holes$\rightarrow$wave pattern is observed. For the sake of learning the wave pattern further, we show the time evolution process of the pattern formation with two special initial conditions and find spiral wave pattern emerge. That is to say, our two-dimensional spatial patterns may indicate the vital role of phase transience regimes in the spatiotemporal organization of the phytoplankton and zooplankton in the aquatic systems. It is also important to distinguish between ``intrinsic" patterns, i.e., patterns arising due to trophic interaction such as those considered above, and ``forced" patterns induced by the inhomogeneity of the environment. The physical nature of the environmental heterogeneity, and thus the value of the dispersion of varying quantities and typical times and lengths, can be essentially different in different cases.

\vspace*{0.5cm}

\section*{Acknowledgements}
This work is supported by University Grants Commission, Govt. of India under grant no. F.33-116/2007(SR) to the corresponding author (RKU) who has visited to Budapest, Hungary under Indo-Hungarian Educational exchange programme. Authors are also grateful to Dr. Sergei V. Petrovskii and the reviewer for their critical review and helpful suggestions.


\section*{Appendix: Proof of Theorem 1}

The Jacobian matrix of model~\eqref{31} corresponding to the positive equilibrium $E^*(P^*, H^*)$ is given by
$$J=\left[\begin{array}{cc}b_{11} &
b_{12} \\
b_{11} & b_{11} \end{array}
\right],$$
where
$$\begin{array}{l}b_{11}=r-2B_1P^*-\frac{2B_2D^2P^{*}H^*}{(P^{*2}+D^2)^2},\qquad
b_{12}=-\frac{B_2P^{*2}}{P^{*2}+D^2}, \\[8pt]
b_{21}=\frac{C_2H^{*2}}{P^{*2}}, \qquad b_{22}=C_1-\frac{2C_2H^{*}}{P^{*}}-\frac{2D_1^2FH^*}{(H^{*2}+D_1^2)^2}.\end{array}$$

Since $E^*(P^*, H^*)$  is a solution of Eq.~\eqref{32}. From the first equation of Eq.~\eqref{32}, we obtain
$$r-B_1P^*=\frac{B_2P^{*}H^*}{P^{*2}+D^2},$$
So,
$$\begin{array}{l}
b_{11}=r-2B_1P^*-\frac{2B_2D^2P^{*}H^*}{(P^{*2}+D^2)^2}=-B_1P^*+(r-B_1P^*)-\frac{2B_2D^2P^{*}H^*}{(P^{*2}+D^2)^2}\\[8pt]
\quad\,\,\, =-B_1P^*+\frac{B_2P^{*}H^*}{P^{*2}+D^2}-\frac{2B_2D^2P^{*}H^*}{(P^{*2}+D^2)^2}
=-B_1P^*+\frac{B_2P^{*}H^*(P^{*2}-D^2)}{(P^{*2}+D^2)^2}\\[8pt]
\quad\,\,\, =P^*(-B_1+B_2H^*f_1(P^*, D))
\end{array}$$
where, $f_1(P^*, D)=\frac{P^{*2}-D^2}{(P^{*2}+D^2)^2}$.

From the second equation of Eq.~\eqref{32}, we have
$$C_1-\frac{C_2H^{*2}}{P^{*2}}=\frac{FH^*}{H^{*2}+D_1^2}.$$
Therefore,
$$\begin{array}{l}
b_{22}=C_1-\frac{2C_2H^{*2}}{P^{*2}}-\frac{2D_1^2FH^*}{(H^{*2}+D_1^2)^2}
=-\frac{C_2H^{*2}}{P^{*2}}+\left(C_1-\frac{C_2H^{*2}}{P^{*2}}\right)-\frac{2D_1^2FH^*}{(H^{*2}+D_1^2)^2}\\[8pt]
\quad\,\,\, =-\frac{C_2H^{*2}}{P^{*2}}+\frac{FH^*}{H^{*2}+D_1^2}-\frac{2D_1^2FH^*}{(H^{*2}+D_1^2)^2}
=-\frac{C_2H^{*2}}{P^{*2}}+\frac{FH^*(H^{*2}-D_1^2)}{(H^{*2}+D_1^2)^2}\\[8pt]
\quad\,\,\, =H^*\left(-\frac{C_2}{P^*}+Ff_2(H^*, D_1)\right),
\end{array}$$
where $f_2(H^*, D_1)=\frac{(H^{*2}-D_1^2)}{(H^{*2}+D_1^2)^2}$.

Now the characteristic equation corresponding to the matrix $J$ is given by
$$\lambda^2-(b_{11}+b_{22})\lambda+(b_{11}b_{22}-b_{12}b_{21})\quad \text{or}\quad \lambda^2+A\lambda+B=0,$$
where $A$ and $B$ are defined in~\eqref{33}. By Routh-Hurwitz criteria, all eigenvalues of $J$ will have negative real parts if and only if $A>0, B>0$. Thus, the theorem follows.

\end{document}